\documentclass[twocolumn,prc,aps,showpacs,superscriptaddress,amsmath,amssymb,floatfix,nofootinbib]{revtex4}
\usepackage{graphicx}

\newcommand{\ratt}{$R_A$}
\newcommand{\rholsq}{$\langle t_I^2 \rangle$}
\newcommand{\rhol}{$\langle t_{II} \rangle$}
\newcommand{\ud}{\mathrm{d}}

\begin{document}

\title{$A$-dependence of hadronization in nuclei}

\author{H.P.~Blok}
\email{henkb@nikhef.nl}
\affiliation{Department~of~Physics~and~Astronomy, Vrije~Universiteit, de~Boelelaan~1081, 1081~HV~Amsterdam, The~Netherlands}
\affiliation{Nationaal Instituut voor Kernfysica en Hoge-Energie Fysica (NIKHEF),
P.O.~Box~41882, 1009~DB~Amsterdam, The~Netherlands}
\author{L.~Lapik\'as}
\email{louk@nikhef.nl}
\affiliation{Nationaal Instituut voor Kernfysica en Hoge-Energie Fysica (NIKHEF),
P.O.~Box~41882, 1009~DB~Amsterdam, The~Netherlands}

\date{\today}

\begin{abstract}
The $A$-dependence of models for the attenuation of hadron
production in semi-inclusive deep-inelastic scattering
on a nucleus is investigated for realistic matter distributions.
It is shown that the dependence for a pure partonic (absorption)
mechanism is more complicated than a simple $A^{2/3}$ ($A^{1/3}$)
behavior, commonly found when using rectangular or Gaussian distributions,
but that the $A$-dependence may still be indicative for
the dominant mechanism of hadronization.
\end{abstract}


\maketitle

The study of hadronization, the process that leads from partons produced
in some elementary interaction to the hadrons observed experimentally,
is of importance, both in its own right as a study of a non-perturbative
QCD process, and in the interpretation
of data from experiments that use outgoing hadrons as a tag.
The end products of the hadronization process in free space are
known from $e^+\,e^-$ annihilation, but very little is known
about the space-time development of the process.
One way to investigate this is to study the semi-inclusive production
of hadrons in deep-inelastic scattering of electrons from a nucleus,
where the nucleus is used as a length(time)-scale probe
(see Refs.~\cite{air01,air03}).

\begin{figure}[!tb]
\includegraphics[width=6cm]{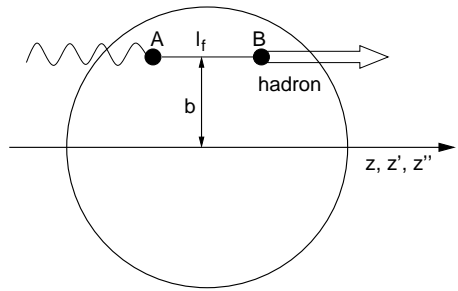}
\caption{\label{fig:hadr_in_nucl}
Illustration of hadronization in a nucleus and used coordinates.
The parton is produced at point A, while the hadron is formed at B.
The distance $l_f$ has a distribution with as average the formation
length $L_f$.}
\end{figure}

Even if hadronization is not yet quantitatively understood, it is known
that the following processes play a role in leptoproduction of hadrons
in a nucleus (see also Fig.~\ref{fig:hadr_in_nucl}).
After a quark in a nucleon is hit by the virtual photon,
it loses energy by scattering from other quarks and radiating gluons,
thus creating quark-antiquark pairs. After some time\footnote{For a more detailed discussion of the concept of formation time etc. see Ref. \cite{bia87}.}
and corresponding length $l_f$
colorless (pre)hadrons\footnote{For the present discussion it is not needed to discriminate between hadrons and prehadrons, so in the remainder we will just talk
about hadrons (but see, e.g., Refs.~\cite{kop04,fal04}).} are formed.
The average value $L_f$ of the formation length has been
estimated~\cite{bia87} based on the Lund model to be typically
in the range of 1--10~fm if the virtual-photon energy is in the
range 5--30~GeV. Hence, $L_f$ is comparable to the size of a nucleus.

If the hadronization is fast, i.e., the hadrons are produced
inside the nucleus, they can be absorbed, which will show up
as an `attenuation' of the hadron yield.
(In experiments one measures the ratio \ratt\ of the yield
on a nucleus with mass number $A$ and the one on a deuteron.)
If, on the other hand, the hadronization is stretched out over
distances large compared to the size of a nucleus, the
relevant interactions will be partonic, involving the emission of gluons
and quark-gluon interactions, which also changes the production
of hadrons.

Since hadronization, as a non-perturbative QCD process, cannot be
calculated from first principles, various models have been developed
(see, e.g., Refs.~\cite{wan01,arl01,acc05,kop04,fal04,acc05b})
to describe hadron production and attenuation in a nucleus.
Some models focus on the partonic part, while
others include or emphasize the hadronic part.
In all cases a sizable dependence on the mass number $A$ is predicted.
However, often the calculations use simple forms for the matter
distribution of the nucleus.

In this Report we investigate the $A$-dependence using realistic matter
distributions. We will do this for two schematic models, covering the
extremes sketched above.
The first one (called model I) assumes a purely partonic mechanism,
hadronization occurring outside of the nucleus
(point B in Fig.~\ref{fig:hadr_in_nucl} effectively at infinity)
and thus absorption of the produced hadrons playing no role.
The nuclear attenuation is then caused by the energy loss of the produced
quark due to multiple scattering and gluon bremsstrahlung,
which gives rise to a change of the hadron fragmentation function
and thus to nuclear attenuation.
Because of the Landau-Pomeranchuk-Migdal interference effect \cite{landau}
this energy loss depends quadratically on the length of matter
traversed by the hit quark (for the present discussion we neglect contributions from other quarks; see Refs. \cite{wan01,arl01} for more details).
In terms of the picture of Fig.~\ref{fig:hadr_in_nucl} it depends on
the square of
the density-averaged distance the parton travels within the nucleus
from the point where it is created (point A in Fig.~\ref{fig:hadr_in_nucl}).

On the other hand our second schematic model (called model II) assumes that a possible
attenuation is completely due to absorption of the produced hadron,
so nothing happens between the time the parton is produced and the
hadron is formed (points A and B in Fig.~\ref{fig:hadr_in_nucl}).
In a Glauber approach this attenuation depends on the cross section for
absorption of the hadron\footnote{In our simple model we do not take into account coupled-channels processes. Possible effects of these are discussed in Ref. \cite{fal04}.}
and the density-averaged distance the hadron travels
through the nucleus after it has been formed (see Fig.~\ref{fig:hadr_in_nucl}).
Taking the absorption cross section to be constant (and small enough that
a linear approximation is sufficient), the latter determines the nuclear
attenuation.

This means we have to calculate the following two integrals:
\begin{eqnarray}
\langle t_I^2\rangle&=&\frac{2 \pi}{A}\int_0^\infty b \ud b
\int_{-\infty}^\infty \ud z \, \rho_A(b,z) \nonumber\\
&&\biggl[\int_z^\infty \ud z' \, \rho_A(b,z') \biggr]^2 , \qquad\qquad {\rm (model \ I)} \\
\langle t_{II} \rangle&=&\frac{2 \pi}{A} \int_0^\infty b \ud b
\int_{-\infty}^\infty \ud z \,\rho_A(b,z) \nonumber\\
&&\int_z^\infty \ud z' \, L_f^{-1} e^{-(z'-z)/L_f} \nonumber\\
&&\int_{z'}^\infty \ud z'' \, \rho_A(b,z'') . \qquad\qquad\quad {\rm (model \ II)}
\end{eqnarray}
\noindent
Here the exponential models the distribution of the formation distances $l_f$,
and the matter densities $\rho_A$ are normalized to $A$.
(By entering these quantities with corresponding dynamical factors
into the appropriate formula's of the models, hadron production cross
sections and from these values for the attenuation \ratt\ can be calculated.
However, here we are interested in the $A$-dependence (and moreover
the used models are extreme and schematic).
Under the assumption that the cross section is linear in \rholsq\ and \rhol,
which is a good first-order approximation, the $A$-dependence of these
quantities carries over into the one of the cross sections.)

It can easily be shown that for a nucleus with a mass density
distribution described by one scale parameter, as in the case
of a rectangular (as in the liquid-drop model) or a Gaussian distribution,
the value of \rholsq\ is proportional to the equivalent radius
or rms radius squared, which in those cases leads to an $A^{2/3}$ dependence.
In case of model II one finds for \rhol\ for a rectangular distribution
an $A^{1/3}$ dependence  when $L_f=0$, and a larger exponent\footnote
{This effect of taking into account a distribution of formation
distances has been noted already in Ref.~\cite{acc05}. }
(e.g., about 0.55 for $L_f=4$ fm) at larger $L_f$,
and similar for a Gaussian distribution.

\begin{figure}[!tb]
\includegraphics[width=7.5cm]{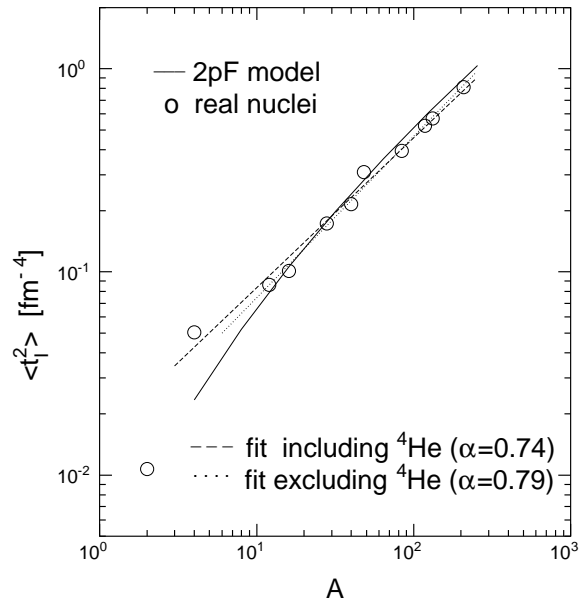}
\caption{\label{fig:rholl}
Values of \rholsq\ as function of $A$ for 2-parameter Fermi (2pF)
matter distributions (solid curve) and for actual nuclei (circles).
The dashed line is a power-law ($A^\alpha$) fit to the latter, excluding
the point for the deuteron. The dotted line in addition excludes the point for $^4$He in the fit.}
\end{figure}

\begin{figure}[tb]
\includegraphics[width=7.5cm]{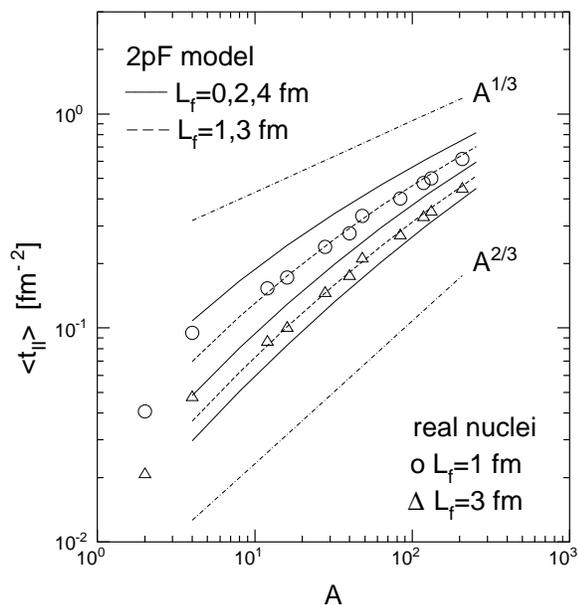}
\caption{\label{fig:rhol1}
Values of \rhol\ as function of $A$ for $L_f=0,2,4$~fm (solid curves, top to bottom)
and $L_f=1,3$~fm (dashed curves, top to bottom)
 for 2pF matter distributions and for actual nuclei (symbols) for
$L_f=1,3$~fm only. The dot-dashed lines represent (arbitrarily normalized) $A^{1/3}$ and $A^{2/3}$ behaviors, as indicated.}
\end{figure}

However, neither a rectangular nor a Gaussian distribution is a good representation
of the mass distribution of a real nucleus.
Therefore we have evaluated \rholsq\ and \rhol\ for a more realistic
distribution, described by a 2-parameter Fermi (Saxon-Woods) form
\begin{equation}
\label{eq:twopf}
\rho_A(r) = \rho_0 / [1+e^{-(r-c)/a}]
\end{equation}
with parameters $\rho_0 = 0.170$ nucleons/fm$^3$, $a=0.5$ fm, and the value
of $c$ so as to give a nucleus with $A$ nucleons.
(This form gives a reasonably good global description of the mass distribution
down to low values of $A$).
The results are given in Table \ref{tab:rhorealisticval} and are shown
in Figs.~\ref{fig:rholl} and \ref{fig:rhol1}.

\begin{table}[!tb]
\caption{\label{tab:rhorealisticval}
Values of \rholsq\ and  \rhol\ for nuclei of mass number A, calculated with
a 2pF distribution (see Eq. \ref{eq:twopf}) with half radius $c$ (second column)
and $\rho_0=0.170$ fm$^{-3}$ and $a=0.5$ fm fixed.
The values
of \rhol\ were calculated for five different values of $L_f$ (columns 4--8).}
\begin{ruledtabular}
\begin{tabular}{r|c|c|ccccc}
 A & c & \rholsq  &  \multicolumn{5}{c}{\rhol}   \\
   & [fm]  &[fm$^{-4}$] &  \multicolumn{5}{c}{[fm$^{-2}$]} \\
\hline
           &       &        &   \multicolumn{5}{c}{$L_f$ [fm]} \\
           &       &        & 0.0    & 1.0    & 2.0 & 3.0 & 4.0 \\
\hline
  4 & 1.321  & 0.023  & 0.108  & 0.070  & 0.048  & 0.037  & 0.030 \\
  8 & 1.875  & 0.052  & 0.166  & 0.113  & 0.081  & 0.062  & 0.051 \\
 16 & 2.531  & 0.106  & 0.243  & 0.175  & 0.129  & 0.101  & 0.083 \\
 32 & 3.324  & 0.200  & 0.342  & 0.260  & 0.199  & 0.159  & 0.132 \\
 64 & 4.295  & 0.359  & 0.468  & 0.373  & 0.295  & 0.242  & 0.204 \\
125 & 5.452  & 0.609  & 0.619  & 0.513  & 0.421  & 0.353  & 0.303 \\
254 & 6.976  & 1.037  & 0.818  & 0.703  & 0.597  & 0.513  & 0.448 \\
\end{tabular}
\end{ruledtabular}
\end{table}

It can be seen from Fig.~\ref{fig:rholl} that for a 2pF matter distribution the A-dependence of \rholsq\ is rather steep.
If one tries to describe it with the power law $A^\alpha$, it would require
a value of $\alpha$ well in excess of 2/3.
Fig.~\ref{fig:rhol1} shows for the case of model II that the slope of the curve, in essence the value of $\alpha$, increases when $L_f$ increases.
Values range from  about $\alpha=0.40$ for $L_f=0$~fm
to $\alpha=0.60$ for $L_f=4$~fm.

Given these findings, and
since it is known that the parameters for actual nuclei are
slightly irregular due to, e.g., shell closures, it is interesting
to see the behavior of \rholsq\ and \rhol\ for real nuclei,
since those are used in experiments.
For that purpose we have used parameterizations \cite{vri87,fri95}
of measured charge distributions (since the neutron distribution
is very similar the error introduced by using the charge distribution
instead of the matter distribution is small, and irrelevant for
the conclusions of the present study)
for the nuclei $^2$H, $^4$He, $^{12}$C, $^{16}$O, $^{28}$Si, $^{40}$Ca,
$^{48}$Ca, $^{84}$Kr, $^{118}$Sn, $^{132}$Xe and $^{208}$Pb.
The results are also given in Table \ref{tab:rhoval} and shown as the symbols
in Figs.~\ref{fig:rholl} and \ref{fig:rhol1}.

\begin{table}[tb]
\caption{\label{tab:rhoval}
Calculated values of \rholsq\ and  \rhol\ for various nuclei. The rms radii
of the employed mass distributions are given in the second column. The values
of \rhol\ were calculated for five different values of $L_f$ (columns 4--8).}
\begin{ruledtabular}
\begin{tabular}{c|c|c|ccccc}
 nucleus & $r_{rms}$ & \rholsq  &  \multicolumn{5}{c}{\rhol}   \\
         & [fm]  &[fm$^{-4}$] &  \multicolumn{5}{c}{[fm$^{-2}$]} \\
\hline
           &       &        &   \multicolumn{5}{c}{$L_f$ [fm]} \\
           &       &        & 0.0    & 1.0    & 2.0 & 3.0 & 4.0 \\
\hline
$^2$H      & 2.13  & 0.011  & 0.068  & 0.041  & 0.027  & 0.021  & 0.017 \\
$^{4}$He   & 1.72  & 0.050  & 0.159  & 0.095  & 0.063  & 0.047  & 0.038 \\
$^{12}$C   & 2.45  & 0.086  & 0.218  & 0.153  & 0.111  & 0.086  & 0.070 \\
$^{16}$O   & 2.73  & 0.101  & 0.238  & 0.172  & 0.127  & 0.100  & 0.082 \\
$^{28}$Si  & 3.08  & 0.173  & 0.316  & 0.239  & 0.181  & 0.145  & 0.120 \\
$^{40}$Ca  & 3.48  & 0.216  & 0.356  & 0.277  & 0.215  & 0.174  & 0.146 \\
$^{48}$Ca  & 3.47  & 0.309  & 0.428  & 0.333  & 0.259  & 0.210  & 0.175 \\
$^{84}$Kr  & 4.25  & 0.395  & 0.493  & 0.403  & 0.325  & 0.270  & 0.230 \\
$^{118}$Sn & 4.67  & 0.523  & 0.571  & 0.475  & 0.390  & 0.328  & 0.281 \\
$^{132}$Xe & 4.83  & 0.570  & 0.598  & 0.500  & 0.413  & 0.348  & 0.300 \\
$^{208}$Pb & 5.51  & 0.811  & 0.719  & 0.615  & 0.519  & 0.445  & 0.387 \\
\end{tabular}
\end{ruledtabular}
\end{table}

\begin{figure}[!tb]
\includegraphics[width=7cm]{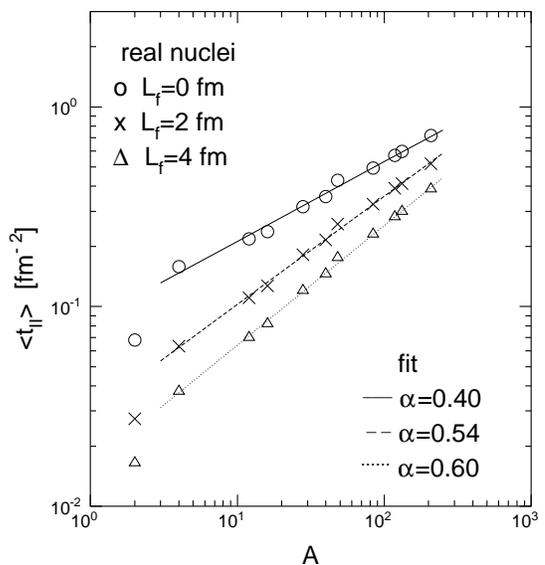}
\caption{\label{fig:rhol2}
Values of \rhol\ as function of $A$ for $L_f=0,2,4$~fm for actual
nuclei (symbols).  The lines are results of power-law ($A^\alpha$) fits, excluding
the point for the deuteron.}
\end{figure}

A first observation is that while the results for $A\geq 4$
scatter around the global curve, the results for $^2$H
are way down. This is due to $^2$H being a rather dilute
system.
Furthermore, it is seen that the results for $^4$He lie
slightly above the global curve, which is related to $^4$He being
a relatively dense system. Fitting the $A$-dependence of the results
for the real nuclei including $^4$He (but leaving out $^2$H),
one finds an exponent of 0.74 for \rholsq\ (see Fig.~\ref{fig:rholl})
and values between 0.40 and 0.60 for \rhol\ depending on the value
of $L_f$ (see fig.~\ref{fig:rhol2}).
When $^4$He is not included in the fit, the values of $\alpha$
in case of \rhol\ change by less than 0.02, whereas
in case of \rholsq\ the value of $\alpha$ increases from 0.74 to 0.79.
Thus, in trying to extract an $A$-dependence from experimental data
it matters if one uses $^4$He as lowest $A$ nucleus, or, $e.g.$, $^{12}$C.

Given these results it would in principle be possible to
discriminate on account of the $A$-dependence between the
two extreme mechanisms that we have used here.
However, in practice the process of hadronization in a nucleus
most probably will be a combination of these mechanisms, with
possibly even different dependences on path lengths in the nucleus
than employed here.
But whatever the mechanism, the $A$-dependence will be an important
ingredient.
Our results show that in model calculations of the $A$-dependence for
comparison with experimental data, global mass distributions, such
as a rectangular distribution employed in the liquid-drop model, are not adequate.
Instead, the use of experimentally established
density functions for the nuclei actually used in the experiment is essential,
where it even matters whether $^4$He with its relatively large density is
included or not.

\begin{acknowledgments}
One of us (HPB) likes to thank Dr. H. J. Pirner for useful discussions.
We thank Dr. J. J. M. Steijger and Dr. G. van der Steenhoven for critically reading the manuscript.
This work is part of the research program of the Stichting voor Fundamenteel
Onderzoek der Materie (FOM).
\end{acknowledgments}

\end{document}